\documentclass[floatfix,aps,prl,superscriptaddress,twocolumn,showpacs]{revtex4-1}

\usepackage{graphicx}
\usepackage{dcolumn}
\usepackage{bm}
\usepackage{amsmath}
\usepackage{color}
\usepackage[colorlinks,
linkcolor=blue,
citecolor=blue,
urlcolor=blue]{hyperref}

\begin{document}

\title{Direction and divergence control of laser-driven energetic proton beam using a disk-solenoid target}

\author{K. Jiang}
\affiliation{Graduate School, China Academy of Engineering Physics, Beijing 100088,
People's Republic of China}
\affiliation{Center for Advanced Material Diagnostic Technology, Shenzhen Technology University, Shenzhen 518118, People's Republic of China}
\author{C. T. Zhou} \email{zcangtao@sztu.edu.cn}
\affiliation{Center for Advanced Material Diagnostic Technology, Shenzhen Technology University, Shenzhen 518118, People's Republic of China}
\affiliation{College of Applied Technology, Shenzhen University, Shenzhen 518060, People's Republic of China}
\affiliation{HEDPS, Center for Applied Physics and Technology and School of Physics, Peking University, Beijing 100871, People's Republic of China}
\author{T. W. Huang}
\affiliation{Center for Advanced Material Diagnostic Technology, Shenzhen Technology University, Shenzhen 518118, People's Republic of China}
\author{L. B. Ju}
\affiliation{Center for Advanced Material Diagnostic Technology, Shenzhen Technology University, Shenzhen 518118, People's Republic of China}
\affiliation{College of Applied Technology, Shenzhen University, Shenzhen 518060, People's Republic of China}
\author{C. N. Wu}
\affiliation{Graduate School, China Academy of Engineering Physics, Beijing 100088, People's Republic of China}
\affiliation{Center for Advanced Material Diagnostic Technology, Shenzhen Technology University, Shenzhen 518118, People's Republic of China}
\author{L. Li}
\affiliation{Center for Advanced Material Diagnostic Technology, Shenzhen Technology University, Shenzhen 518118, People's Republic of China}
\affiliation{HEDPS, Center for Applied Physics and Technology and School of Physics, Peking University, Beijing 100871, People's Republic of China}
\author{H. Zhang}
\affiliation{Center for Advanced Material Diagnostic Technology, Shenzhen Technology University, Shenzhen 518118, People's Republic of China}
\author{S. Z. Wu}
\affiliation{Center for Advanced Material Diagnostic Technology, Shenzhen Technology University, Shenzhen 518118, People's Republic of China}
\author{T. X. Cai}
\affiliation{Center for Advanced Material Diagnostic Technology, Shenzhen Technology University, Shenzhen 518118, People's Republic of China}
\author{B. Qiao}
\affiliation{Center for Advanced Material Diagnostic Technology, Shenzhen Technology University, Shenzhen 518118, People's Republic of China}
\affiliation{HEDPS, Center for Applied Physics and Technology and School of Physics, Peking University, Beijing 100871, People's Republic of China}
\author{M. Y. Yu}
\affiliation{Center for Advanced Material Diagnostic Technology, Shenzhen Technology University, Shenzhen 518118, People's Republic of China}
\author{S. C. Ruan}
\affiliation{Center for Advanced Material Diagnostic Technology, Shenzhen Technology University, Shenzhen 518118, People's Republic of China}
\affiliation{College of Applied Technology, Shenzhen University, Shenzhen 518060, People's Republic of China}

\date{\today}

\begin{abstract}
A scheme for controlling the direction of energetic proton beam driven by intense laser pulse is proposed. Simulations show that a precisely directed and collimated proton bunch can be produced by a sub-picosecond laser pulse interacting with a target consisting of a thin solid-density disk foil with a solenoid coil attached to its back at the desired angle. It is found that two partially overlapping sheath fields are induced. As a result, the accelerated protons are directed parallel to the axis of the solenoid, and their spread angle is also reduced by the overlapping sheath fields. The proton properties can thus be controlled by manipulating the solenoid parameters. Such highly directional and collimated energetic protons are useful in the high-energy-density as well as medical sciences.
\end{abstract}

\maketitle

Energetic laser-driven ion source with unique features, such as small device size and high brightness, is useful in radiography \cite{kug}, warm-dense-matter generation \cite{pel}, fast ignition of fusion core \cite{rot}, isotope generation \cite{led}, tumor therapy \cite{goi}, brightness enhancement for conventional accelerators \cite{kru}, etc. The target-normal sheath acceleration (TNSA) mechanism \cite{wil} is widely investigated because of its undemanding laser and target parameter requirements \cite{mak,cla,bor,fuc,rob,wag}. In TNSA, an intense laser interacts with a target, generating hot relativistic electrons that penetrate through the latter and establish in the backside vacuum a TV/m sheath electric field, which  accelerates the target-back ions to multi-MeV energies \cite{sna}. However, the intrinsic target-normal direction and divergence of the TNSA protons limit their applications. Different methods have been proposed for collimation and manipulation of the TNSA protons, including the use of structured targets \cite{son,heg,sch,kar1,schn,sok,bur,pat,cow,che,bar,qia}, electrostatic lens \cite{ton,kar}, laser prepulse \cite{lin}, oblique laser incidence \cite{zho,mor,zei}, etc., but simultaneously controlling the direction and the divergence angle of the TNSA protons remains difficult.

In this Letter, we propose to attach a solenoid to the back of a thin disk target to collimate and guide the TNSA protons. Three dimensional (3D) particle-in-cell (PIC) simulations show that a precisely directed proton bunch with small divergence angle can be obtained. The result is attributed to a uniquely structured sheath field created by the hot electrons from the foil, and is therefore quite different from the electromagnetic-pulse-induced field usually observed in the laser-target-interaction \cite{kar1}. The results agree quite well with that from an analytical model, which is also useful for tailoring the solenoid parameters in order to produce well directed and collimated high-energy proton bunches under given laser and target conditions.

\begin{figure}
\centering
\includegraphics[width=8.5cm]{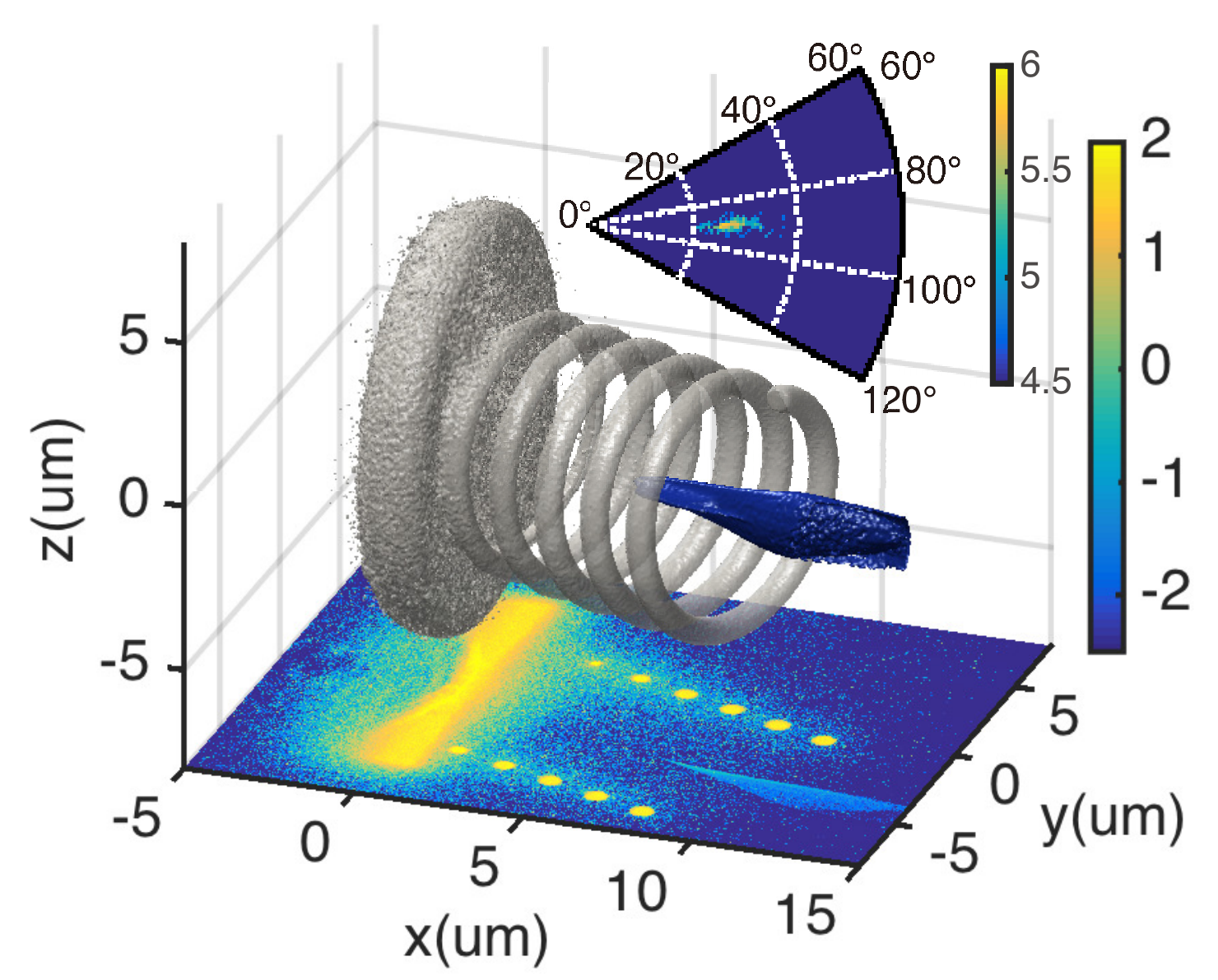}
\caption{The target structure and density isosurfaces from the 3D PIC simulations at $t=420$ fs. The 3D electron and proton density isosurfaces are shown in gray and blue, respectively. The isovalues of the foil and solenoid electrons are $6.12n_c$ and $0.35n_c$, respectively, and the isovalue of the protons (from the foil backside dot) is $0.01n_c$. The projection of the $z=0$ plane at the bottom shows the target electron density $\log_{10}(n_e/n_c)$ (mainly yellow) and the proton density $\log_{10}(n_p/n_c)$ (the well-defined light blue patch). The inset shows the angular density distribution of $>12$ MeV protons in the relevant segment of the proton velocity space $\bm{v}_p$, where the ``angle" is defined by $\arctan(v_{pz}/\sqrt{v_{px}^2+v_{py}^2})$ and the ``radius" by $\arctan(-v_{py}/v_{px})$. \label{fig_1}}
\end{figure}

The target configuration can be visualized in Fig. \ref{fig_1} for the electron and proton densities at $t=420$ fs. The helical plasma-wire solenoid is attached to the back of the disk foil at, for definitiveness, a $\psi=20^{\circ}$ angle. The radius, length, and number of turns of the solenoid are $r_s=3.5 \mu$m, $h=10 \mu$m, and $n=6$, respectively. The coil wire is of diameter $0.6 \mu$m and total length $l=n\sqrt{(2\pi r_s)^2+({h/n})^2}=132 \mu$m. 
The radius and thickness of the foil are $6 \mu$m and $1 \mu$m, respectively. As proton source, a hydrogen dot of thickness $0.5 \mu$m and diameter $1 \mu$m is placed at the center of the rear foil surface. In the 3D PIC simulations with EPOCH \cite{arb}, the foil and the solenoid are Cu$^{2+}$ plasma, at densities $n_{i0}=20n_c$ and $n_{e0}=40n_c$, respectively. The density of the hydrogen dot is $1n_c$, where $n_c \sim 1.1 \times 10^{21} {\rm cm}^{-3}/\lambda_L^2$ is the critical density, and $\lambda_L$ [$\mu$m] is the incident laser wavelength. To account for the laser prepulse, a $5 \mu$m long preplasma of density $n_e=n_{e0}\exp(x/\delta)$, where $\delta=0.5 \mu$m, is placed in front of the disk foil.
A $y$ polarized Gaussian laser pulse of $\lambda_L=1.06 \mu$m, intensity $2 \times 10^{20}$ W/cm$^2$, waist radius $3 \mu$m, and duration $500$ fs is normally incident from the left boundary at $x=-8 \mu$m. The laser has a flat-top temporal profile, with $3.5$ fs rising time. The simulation box is $23 \times 16 \times 16 \mu$m$^3$, with $1143 \times 795 \times 795$ grids. There are 7 macroparticles per cell for the solenoid target and 30 for the hydrogen dot. As shown in Fig. \ref{fig_1}, at $t=420$ fs, a directed and well collimated proton bunch with cut-off energy $>20$ MeV is generated. The exit angle of protons with energy $>12$ MeV is around $26^{\circ}$ from the foil normal in the $(x, y)$ plane.

\begin{figure}
\centering
\includegraphics[width=8.5cm]{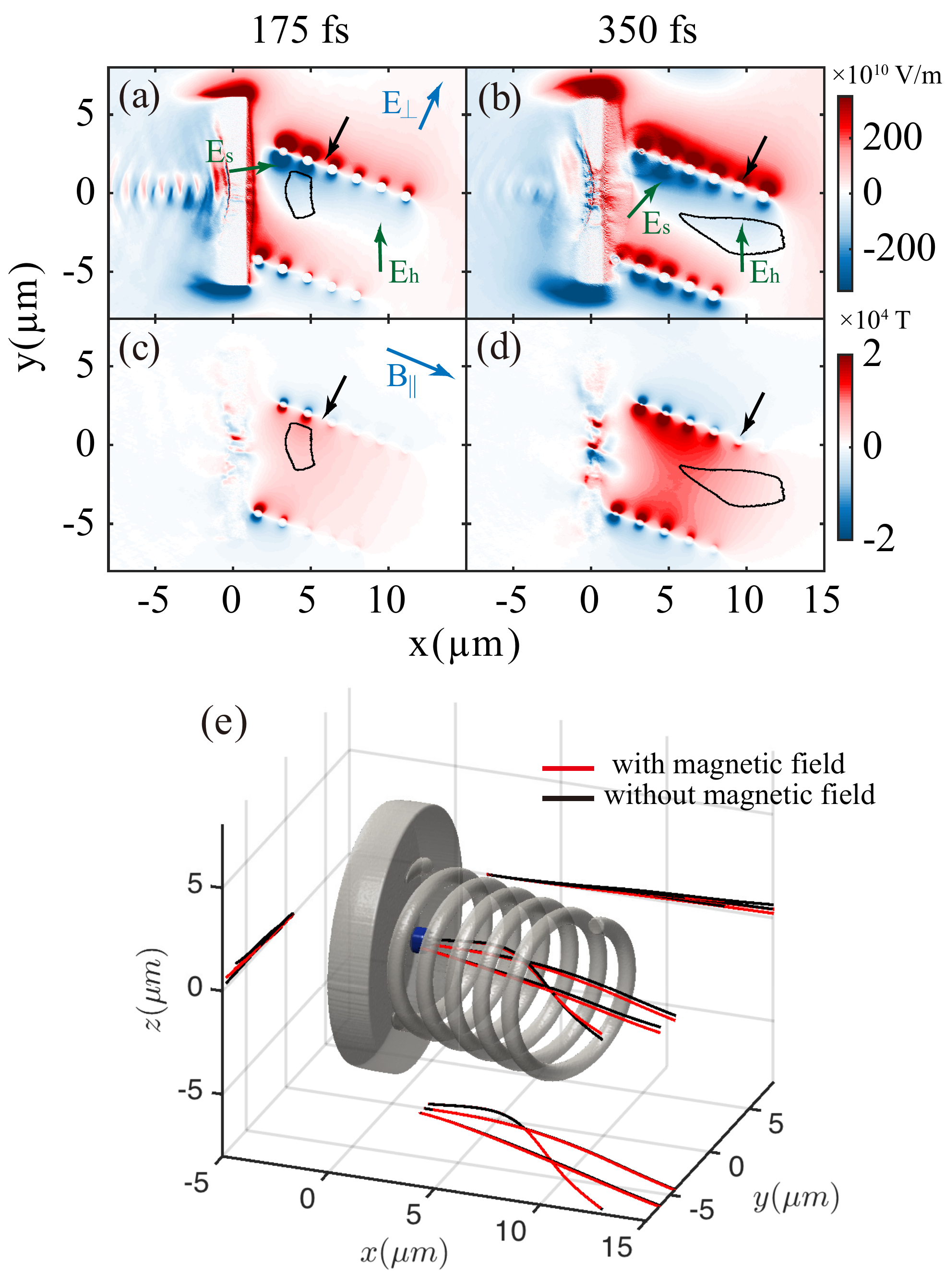}
\caption{Electric and magnetic field profiles on the $z=0$ plane at $t=175$ fs and $t= 350$ fs. (a) and (b) $E_{\perp}$ [V/m], and (c) and (d) $B_{\parallel}$ [T]. The blue arrows in (a) and (c) show the positive directions of $E_{\perp}$ and $B_{\parallel}$, respectively. The black loops roughly outline the TNSA proton bunch, and the black arrows show the leading front of the propagating electrons in the solenoid wire. The green arrows in (a) and (b) show the regions dominated by the sheath fields $E_h$ and $E_s$ generated by the hot sheath electrons and the propagating solenoid electrons, respectively. (e) Several proton trajectories and their projections for with/without the self-generated magnetic field (red/black). The foil-back hot electrons from the target front are relatively homogeneous and not shown for clarity. One can clearly see that the solenoid can effectively guide and collimate the TNSA protons. \label{fig_2}}
\end{figure}

As the intense laser pulse impinges on the target, hot electrons are generated at the front surface and directly accelerated by the laser to high speeds \cite{wik,she}. They can easily penetrate through the foil and establish immediately behind it as well as around the surface of the solenoid wire an intense sheath electric field, roughly given by $E_h\sim T_{eh}/(er_s)\sim 0.86 \times 10^{12}$ V/m, where $T_{eh}$ is the temperature of hot electrons and $e$ is the elementary charge \cite{wil}.
This value agrees well with that from the simulation, as shown in Fig. \ref{fig_2}(a). On the other hand, the electrons exiting the foil at where the solenoid is attached can propagate in the solenoid wire (a conductor) and create around its surface a local sheath field with magnitude $E_t \sim T_{et}/(eK) \approx 2.4\times 10^{12}$ V/m, where $T_{et}$ is the temperature of these secondary electrons and $K$ the spatial scale of their field. Thus in the overlapped region the sheath field on the solenoid surface can be as high as $E_s=E_h+E_t\approx 3.3 \times 10^{12}$ V/m. The normal (to the solenoid surface) components of both $E_{h}$ and $E_{s}$ lead to a focusing force on the dot protons accelerated by the foil-back sheath field, as shown in Figs. \ref{fig_2}(a) and (b), so that instead of propagating normal to the foil back-surface, the protons are directed and collimated by the solenoid, as can be seen in Fig. \ref{fig_2}(b).

It is also necessary to consider the effect of the self-generated magnetic fields. The electron current in the solenoid wire is much larger than the Alfv\'en limit \cite{alf}. The return current induced on the wire surface gives rise to a strong longitudinal magnetic field in the solenoid \cite{kay,xia}. Figures \ref{fig_2}(c) and (d) show that the peak magnetic field (on the solenoid axis) is $B_\parallel\sim 1\times 10^4$ T. The corresponding proton gyroradius is $r_{pg}\sim m_pv_{p_\perp}/(eB_\parallel)\approx 15.7 \mu$m, where $m_p$ and $v_{p_\perp}\sim 0.05c$ are the proton rest mass and transverse velocity component, respectively. Since $r_{pg}\gg r_s$, the magnetic field should have little effect on the protons, as can be seen in Fig. \ref{fig_2}(e) for three typical proton trajectories. Thus, the direction and collimation of the proton bunch are mainly determined by the electric fields $E_{h_\perp}$ and $E_{s_\perp}$.

It is thus of interest to see how $E_{h_{\perp}}$ and $E_{s_{\perp}}$ affect the proton dynamics. Since the solenoid electrons propagate at near light speed, the distance between the field boundary (i.e., the rough boundary between $E_{h_\perp}$ and $E_{s_\perp}$) and the foil back-surface at $t>T_0$ is $L_{b_{\parallel}}(t)\sim ch(t-T_0)/l$, where $T_0$ is the time when the electrons start to propagate in the solenoid wire. The protons are first accelerated to a velocity $\bm{v}_{p_{0}}$ by the sheath field at the foil rear from $T_0$ to $(T_0+t_{\mathrm{acc}})$, after which they are no longer accelerated. The parallel and transverse proton velocities are then $v_{p_{\parallel_0}}=v_{p_0}\cos(\phi +\psi)={\rm const.}$ and $v_{p_{\perp_0}}=v_{p_0}\sin(\phi+\psi)$, respectively, where $\phi$ is the divergence angle of the protons and $\psi$ is the angle between the solenoid and the foil. Since protons with $v_{p_{\parallel_0}}\le {ch}/({l-ct_{\rm acc}})$ cannot cross the field boundary, the protons (hereafter referred to as proton 1) is thus mainly governed by $E_{s_\perp}$ (or $E_{h_\perp}$ for those with small $v_{p_\perp}$). On the other hand, protons (hereafter referred to as proton 2) with $v_{p_{\parallel_0}}>ch/(l-ct_{\rm acc})$ can cross the field boundary at $t=T_1=T_0+t_{\rm acc}/(1-ch/(lv_{p_{\parallel_0}}))$. Type 1 protons satisfy $d^2L_\perp/{dt^2}=eE_{s_\perp}/{m_p}$, where $L_{\perp}$ is the distance between the proton and the solenoid axis. Assuming $E_{s_\perp}$ depends linearly on $L_\perp$, say, $E_{s_\perp}=-E_{s_{\perp},\max}L_{\perp}/r_s$, we obtain for $t>T_0+t_{\rm acc}$
\begin{equation}\label{eq1}
L_{\perp}(t)=\frac{v_{p_{\perp_0}}}{\omega_s}\sin\left\{\omega_s \left[t-(T_0+t_{\rm acc})\right]\right\},
\end{equation}
and
\begin{equation}\label{eq2}
v_{p_\perp}(t)= v_{p_{\perp_0}}\cos\left\{\omega_s \left[t-(T_0+t_{\rm acc})\right]\right\},
\end{equation}
for proton 1, where $\omega_s=\sqrt{eE_{s_{\perp},\max}/(m_pr_s)}$. Therefore, the exit angle for proton 1 is centered along the $\psi$ direction, with the divergence angle given by
\begin{equation}\label{eq3}
|v_{p_\perp}/v_{p_\parallel}|\le\tan(\phi+\psi).
\end{equation}
The dynamics of proton 2 for $t\le T_1$ is similar to that of proton 1. However, at $t=T_1$, proton 2 can cross the field boundary. Thereafter their motion is governed by $d^2L_\perp/dt^2=eE_{h_\perp}/m_p$. Assuming again that $E_{h_\perp}$ depends linearly on $L_\perp$, i.e., $E_{h_\perp}=-E_{h_{\perp},\max}L_\perp/r_s$, one obtains for proton 2
\begin{equation}\label{eq4}
L_{\perp}(t)=c_1\cos[\omega_h(t-T_1)]+c_2\sin [\omega_h(t-T_1)],
\end{equation}
\begin{equation}\label{eq5}
v_{p_\perp}(t)=-c_1\omega_h\sin[\omega_h(t-T_1)] +c_2\omega_h\cos[\omega_h(t-T_1)],
\end{equation}
where $c_1=v_{p_{\perp_0}}\sin(\omega_s\tau)/\omega_s$, $c_2=v_{p_{\perp_0}}\cos(\omega_s\tau)/\omega_h$, $\tau=cht_{\rm acc}/(lv_{p_{\parallel_0}}-ch)$, and $\omega_h=\sqrt{eE_{h_{\perp},\max}/(m_pr_s)}$. Accordingly, the exit angle of the type 2 protons for $t>T_1$ is centered at $\psi$, with the divergence angle given by
\begin{equation}
|v_{p_\perp}/v_{p_\parallel}|\le f(\eta)\tan(\phi+\psi),   \label{eq6}
\end{equation}
where $f(\eta)=\sqrt{\left(E_{h_{\perp},\max}/E_{s_{\perp},\max}\right)\sin^2 \eta+\cos^2\eta}$ and $\eta=\omega_s\tau$. Since $f(\eta)\le 1$, the divergence of the type 2 protons is reduced after they cross the field boundary.

\begin{figure}
\centering
\includegraphics[width=8.5cm]{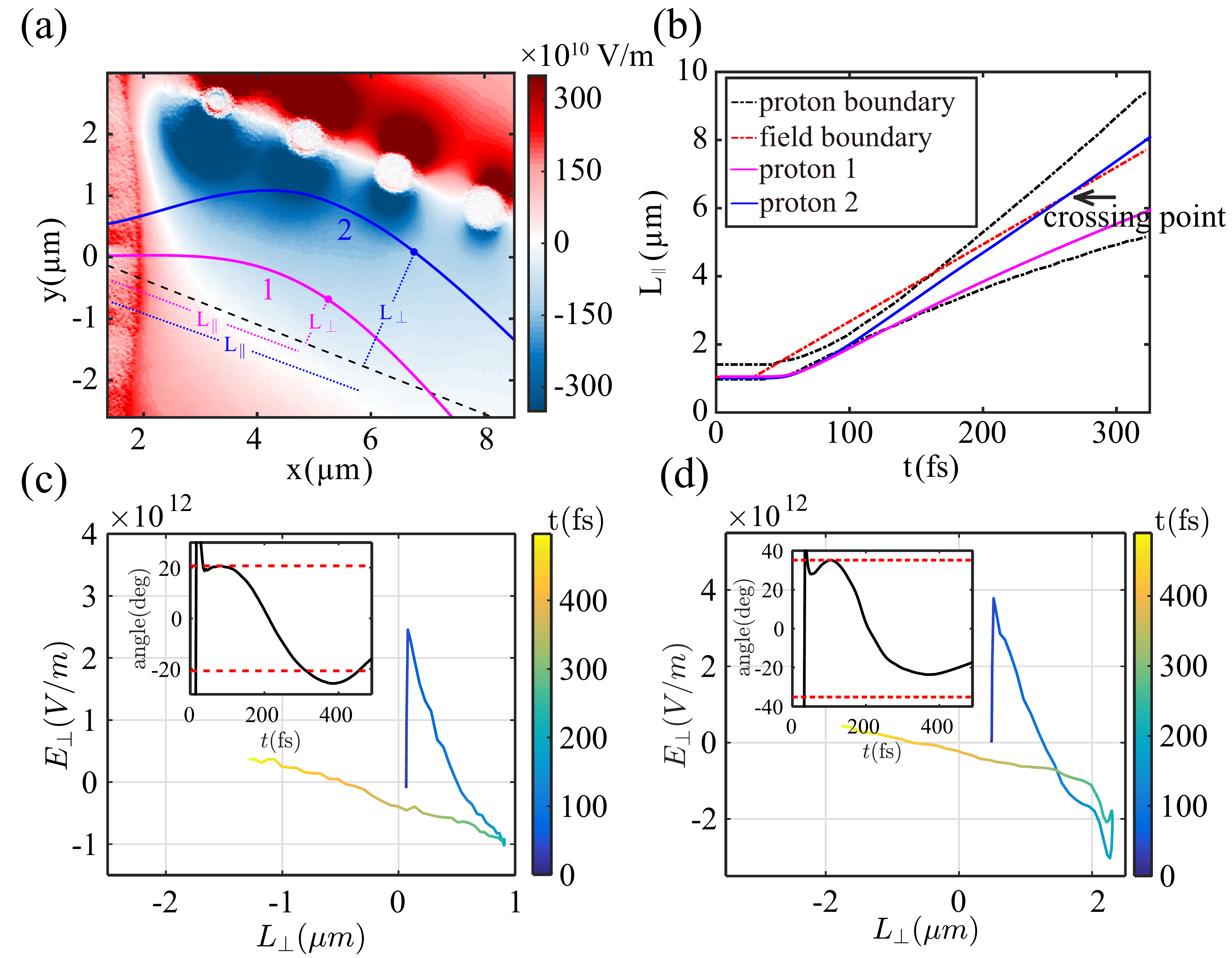}
\caption{(a) Local $E_{\perp}$ [V/m] at $t=259$ fs and the trajectories of protons 1 and 2 (magenta and blue curves, respectively). The dots on the trajectories mark their positions at $t=259$ fs. The black dashed line indicates the axis of the solenoid. (b) Evolution of $L_{\parallel}$ of the protons. (c) $E_{\perp}$ versus $L_{\perp}$ as experienced by proton 1 at different times (see the color bar). The inset shows the evolution of the divergence angle ${\rm arctan}(-v_{p_\perp}/v_{p_\parallel})$ (black curve) with respect to the solenoid axis. (d) Same as (c), but for proton 2. \label{fig_3}}
\end{figure}

Figure \ref{fig_3}(a) shows the trajectories of the two proton types. Although $E_{\perp}$ also evolves with time, only its distribution at $t=259$ fs is displayed in the background as a reference. Note that at this moment proton 2 is located at the field boundary, agreeing well with the boundary crossing time from the theory and shown in Fig. \ref{fig_3}(b). We see that, except immediately behind the foil disk, proton 1 experiences negative $E_{h_\perp}$ at all times. On the other hand, proton 2 is decelerated by $E_{s_\perp}$ when it moves away from the solenoid center axis and accelerated by $E_{h_\perp}$ when it moves toward it.

Figures \ref{fig_3}(c) and (d) show the trajectories of proton 1 and proton 2 in the $E_{\perp}$ versus $L_{\perp}$ space. In (c) we can see that proton 1 experiences a larger $E_{\perp}$ field on the way back to the solenoid axis than that when it moves away from the latter, resulting in increase of its divergence angle. In contrast, in (d) we see that proton 2 experiences a smaller $E_{\perp}$ on the way back to the axis than that when it moves away from it, so that its divergence angle decreases with time. Accordingly, the divergence angle of the TNSA proton bunch can be minimized by tailoring the solenoid parameters such that a maximum number of protons can cross the boundary between the two sheath fields.

\begin{figure}
\centering
\includegraphics[width=8.5cm]{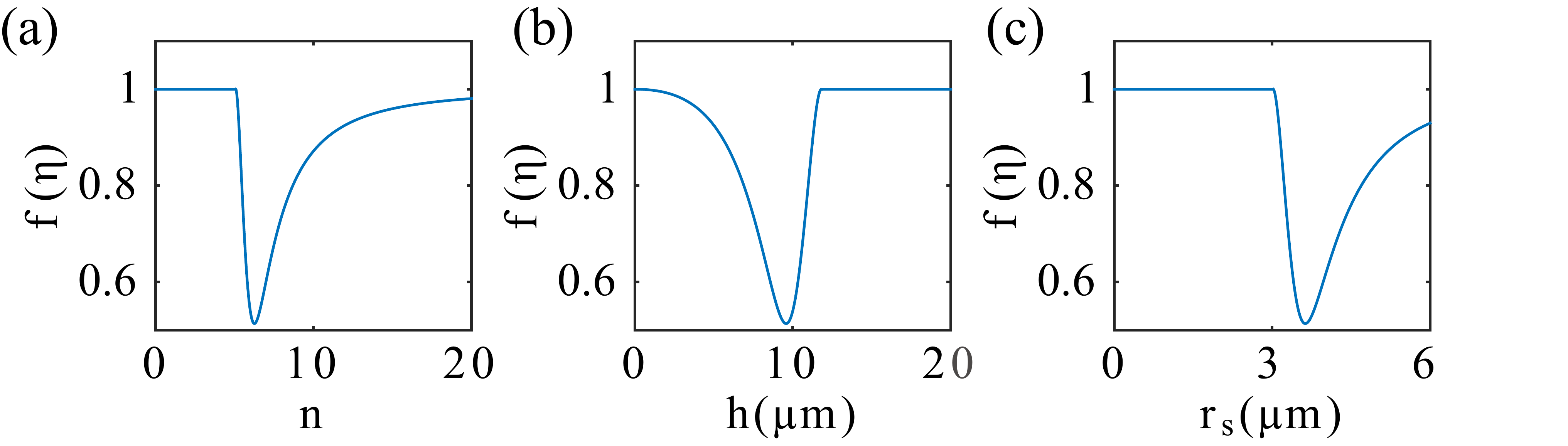}
\caption{Dependence of $f(\eta)$ on
(a) $n$ for $h=10 \mu$m and $r_s=3.5 \mu$m,
(b) $h$ for $n=6$ and $r_s=3.5 \mu$m,
(c) $r_s$ for $n=6$ and $h=10 \mu$m.
Here, $E_{h_{\perp},\max}\approx 0.86\times 10^{12}$ V/m, $E_{s_{\perp},\max}\approx 3.26\times 10^{12}$ V/m, $\phi=5^{\circ}$, $\psi=20^{\circ}$, and $t_{\rm acc}=100$ fs, respectively. \label{fig_4}   }
\end{figure}

\begin{figure}
\centering
\includegraphics[width=8.5cm]{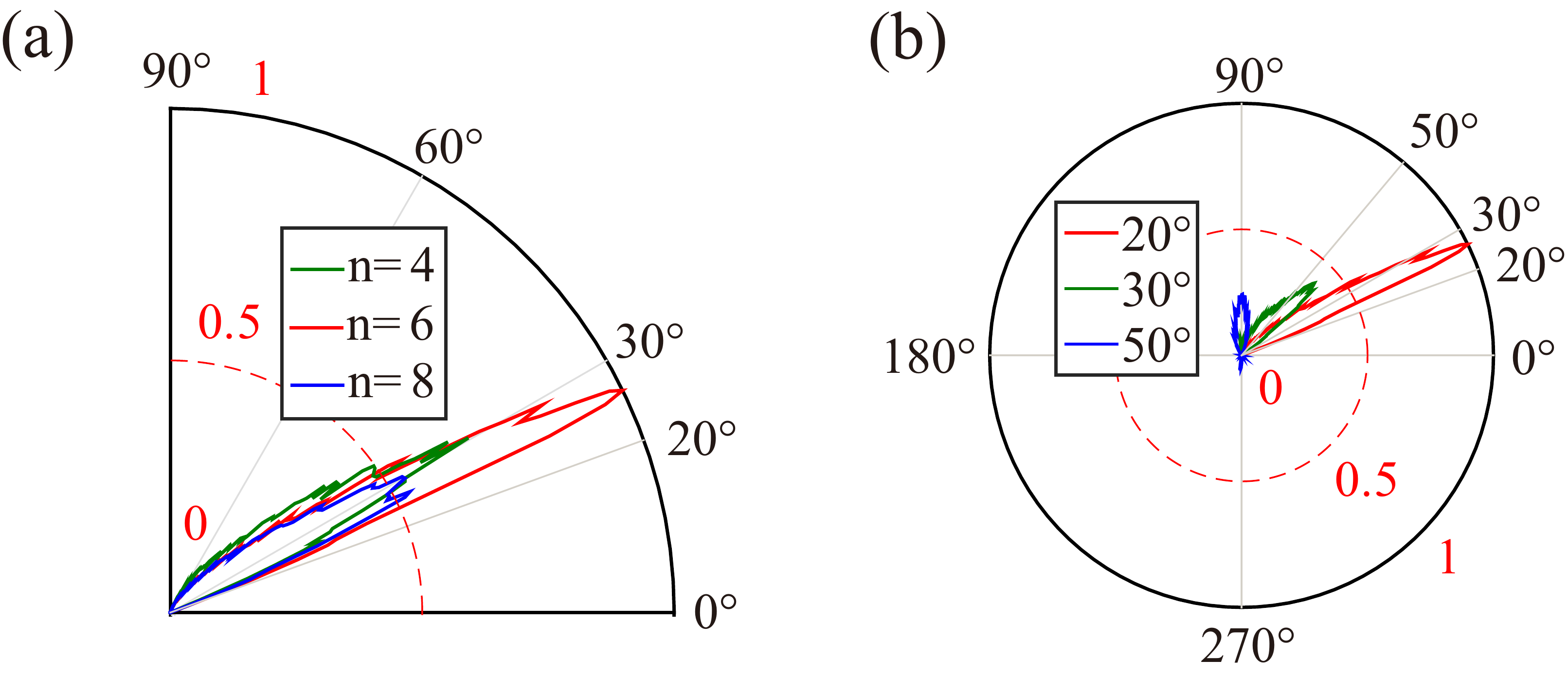}
\caption{Angular distribution of the normalized proton energy density for (a) $\psi=20^{\circ}$, $r_s=3.5 \mu$m, $h=10 \mu$m, and $n=4$, 6, and 8, (b) $\psi=20^{\circ}$, $r_s=3.5 \mu$m, $h=10 \mu$m, and $n=6$ (red curve); $\psi=30^{\circ}$, $r_s=3.5 \mu$m, $h=10 \mu$m, and $n=7$ (green curve); and $\psi=50^{\circ}$, $r_s=3.5 \mu$m, $h=10 \mu$m, and $n=9$ (blue curve). The solenoid parameters here have been optimized for each $\psi$.\label{fig_5}}
\end{figure}

The parameter $f(\eta)$ depends strongly on the solenoid parameters $h$, $n$, and $r_s$. If two of them are fixed, one can find the value of the third one in order to obtain the highest proton energy density, as shown in Fig. \ref{fig_4}. For example, $n=6$ should be the optimal number of turns if $h=10 \mu$m and $r_s=3.5 \mu$m. Indeed, Fig. \ref{fig_5}(a) from the simulation shows that with $n=6$, the resulting proton energy density is higher than that with $n=4$ and $n=8$. Figure. \ref{fig_5}(b) for the exit angle of the protons versus the solenoid angle $\psi$ under optimized conditions shows that the direction of the proton bunch is well controlled if $\psi\le 20^{\circ}$. However, both the directional preciseness and proton energy density decrease with increase of $\psi$, which is also consistent with the relations (\ref{eq3}) and (\ref{eq6}) of the analytical model.

In summary, we have proposed an effective scheme for directional control and collimation of intense laser-driven protons using a disk-solenoid target. Our simulations show that two partially overlapping sheath fields are induced by the hot electrons and they result in an electric field distribution that collimates and focuses the energetic protons in the solenoid to its axis (instead of the target normal direction). In fact, the divergence angle of the protons decreases when they cross the boundary region of the two sheath fields. The simulation results are in good agreement with that of an analytical model of the proton dynamics, which is also useful for tailoring the solenoid parameters for obtaining the desired proton energy and divergence angle. Highly collimated and precisely directed proton bunches are desirable for radiography, tumor therapy, warm dense matter generation, etc.

This work is supported by the National Key Program for R\&D Research and Development, Grant No. 2016YFA0401100; the National Natural Science Foundation of China (NNSFC), Grant Nos.  11575031, 11575298, and 11705120, the China Postdoctoral Science Foundation 2017M612708, the National ICF Committee of China. K. J. would like to thank K. D. Xiao, C. Z. Xiao, R. Li, T. Y. Long, Y. C. Yang, R. X. Bai, and M. J. Jiang for useful discussions and help.


\end{document}